# Demonstration of Enhanced Monte Carlo Computation of the Fisher Information for Complex Problems

Xumeng Cao

3400 North Charles Street, AMS Department

Johns Hopkins University

Baltimore, Maryland 21218, USA

**ABSTRACT**

The Fisher information matrix summarizes the amount of information in a set of data relative to the quantities of interest. There are many applications of the information matrix in statistical modeling, system identification and parameter estimation. This short paper reviews a feedback-based method and an independent perturbation approach for computing the information matrix for complex problems, where a closed form of the information matrix is not achievable. We show through numerical examples how these methods improve the accuracy of the estimate of the information matrix compared to the basic resampling-based approach. Some relevant theory is summarized.

# 1. INTRODUCTION

The Fisher information matrix plays an essential role in statistical modeling, system identification and parameter estimation [see Ljung 1999 and Bickel and Doksum 2007, Section 3.4]. Consider a collection of $n$ random vectors $Z = [z_1, z_2, \ldots, z_n]^T$, where each $z_i$ is a vector for $i = 1, 2, \ldots, n$. These vectors are not necessarily independent and identically distributed. Let us assume that the probability density/mass function for $Z$ is $p_Z(\zeta|\theta)$, where $\zeta$ is a dummy matrix representing a possible realization of $Z$; $\theta \in \Theta$ is the unknown $p \times 1$ parameter vector. The corresponding likelihood function is

$$l(\theta|\zeta) = p_Z(\zeta|\theta).$$

Letting $L(\theta) = -\log l(\theta|Z)$ be the negative log-likelihood function, the $p \times p$ Fisher information matrix $F(\theta)$ for a differentiable $L$ is given by

$$F(\theta) \equiv E\left(\frac{\partial L}{\partial \theta} \frac{\partial L}{\partial \theta^T}\right), \tag{1}$$

where the expectation is taken with respect to the data set $Z$.

Except for relatively simple problems, however, the definition of $F(\theta)$ in (1) is generally not useful in practical calculation of the information matrix. Computing the expectation of a product of multivariate nonlinear functions is usually a formidable task. A well-known equivalent form follows from the assumption that $L$ is twice continuously differentiable in $\theta$. That is, the Hessian matrix

$$H(\boldsymbol{\theta}) = \frac{\partial^2 L}{\partial \boldsymbol{\theta} \partial \boldsymbol{\theta}^T}$$

is assumed to exist. Furthermore, assume that $l$ is regular in the sense that standard conditions such as in Wilks (1962, pp. 408–411 and 418–419) or Bickel and Doksum (2007, p. 179) hold. Specifically, $l$ satisfies the following regularity conditions:

(I) The support of $l$: $\mathbf{A} = \{\zeta: l(\boldsymbol{\theta}|\zeta) > 0\}$ does not depend on $\boldsymbol{\theta}$. For all $\zeta \in \mathbf{A}$, $\boldsymbol{\theta} \in \Theta$, $\partial L/\partial \boldsymbol{\theta}$ exists and is finite;

(II) If $T(\zeta)$ is any statistic such that $E_{\boldsymbol{\theta}}(|T(\zeta)|) < \infty$ for all $\boldsymbol{\theta} \in \Theta$, the integrals

$$\int T(\zeta) \frac{\partial l(\boldsymbol{\theta}|\zeta)}{\partial \boldsymbol{\theta}} d\zeta \quad \text{and} \quad \int \left| T(\zeta) \frac{\partial l(\boldsymbol{\theta}|\zeta)}{\partial \boldsymbol{\theta}} \right| d\zeta$$

are continuous functions of $\boldsymbol{\theta}$. Under such conditions, the information matrix is related to the Hessian matrix of $L$ through:

$$\boldsymbol{F}(\boldsymbol{\theta}) = E\left( \frac{\partial^2 L}{\partial \boldsymbol{\theta} \partial \boldsymbol{\theta}^T} \right). \tag{2}$$

The form of $\boldsymbol{F}(\boldsymbol{\theta})$ in (2) is usually more amenable to calculate than the product-based form in (1).

In many practical problems, however, closed forms of $\boldsymbol{F}(\boldsymbol{\theta})$ do not exist. In such cases, we need to estimate the Fisher information (see Al-Hussaini and Ahmad 1984, Mainassara et al. 2011, Gibson and

Ninness 2005, and Spall and Garner 1990). Given the equivalent form of $F(\theta)$ in (2), we can estimate $F(\theta)$ using measurements of $H(\theta)$. The conventional approach uses resampling-based method to approximate $F(\theta)$. In this paper, we demonstrate two other enhanced Monte Carlo methods: feedback-based approach and independent perturbation approach; see Spall (2008). The Monte Carlo computation of $F(\theta)$ is discussed in other scenarios too, see Das et al. (2010), where prior information of $F(\theta)$ is used in estimation. The remainder of the paper is organized as follows: in Section 2, we introduce methodology of three different approaches discussed in this paper; some relevant theory is summarized in Section 3; section 4 includes two numerical examples and discussions on relative performance of the three methods; a brief conclusion is made in section 5.

## 2. METHODOLOGY

### 2.1. Basic Resampling-based Approach

We first give a brief review of a Monte Carlo resampling-based approach to compute $F(\theta)$, as given in Spall (2005). Let $Z_{\text{pseudo}}(i)$ be a collection of Monte Carlo generated random vectors from the assumed distribution based on the parameters $\theta$. Letting $\hat{H}_{k|i}$ represent the $k$th estimate of $H(\theta)$ at the data set $Z_{\text{pseudo}}(i)$, we generate $\hat{H}_{k|i}$ via efficient simultaneous perturbation (SPSA) principles:

$$\hat{H}_{k|i} = \frac{1}{2}\left[\frac{\delta g_{k|i}}{2c}(\Delta_{k|i}^{-1})^T + \left(\frac{\delta g_{k|i}}{2c}(\Delta_{k|i}^{-1})^T\right)^T\right], \tag{3}$$

where $\delta g_{k|i} = g(\theta + c\Delta_{k|i} | Z_{\text{pseudo}}(i)) - g(\theta - c\Delta_{k|i} | Z_{\text{pseudo}}(i))$, $g(\cdot)$ is the exact or estimated gradient function of $L$, depending on the information available; $\Delta_{k|i} = [\Delta_{k1|i}, \Delta_{k2|i}, ..., \Delta_{kp|i}]^T$ is a mean-zero random vector such that the scalar elements are i.i.d. symmetrically distributed random variables that are uniformly bounded and satisfy $E(|1/\Delta_{kj|i}|) < \infty$, $\Delta_{k|i}^{-1}$ denotes the vector of inverses of the $p$ individual elements of $\Delta_{k|i}$, and $c > 0$ is a "small" constant.

The Monte Carlo approach of Spall (2005) is based on a double averaging scheme. The first "inner" average forms Hessian estimates at a given $Z_{\text{pseudo}}(i)$ ($i = 1, 2, ..., N$) from $k = 1, 2, ..., M$ values of $\hat{H}_{k|i}$ and the second "outer" average combines these sample mean Hessian estimates across the $N$ values of pseudo data. Therefore, the "basic" Monte Carlo resampling-based estimate of $F(\theta)$ in Spall (2005), denoted as $\bar{F}_{M,N}(\theta)$, is:

$$\bar{F}_{M,N}(\theta) = \frac{1}{N} \sum_{i=1}^{N} \frac{1}{M} \sum_{k=1}^{M} \hat{H}_{k|i}.$$

This resampling-based estimation method is easy to implement and works well in practice. However, this basic Monte Carlo approach could be improved by some extra effort. In the next two subsections, we introduce the use of feedback information and independent perturbation, respectively.

**2.2 Enhancements Through Use of Feedback**

The feedback ideas for FIM estimation in Spall (2008) are related to the feedback ideas presented with the most updates in Spall (2009), as applied to stochastic approximation. From Spall (2009), it is known that $\hat{H}_{k|i}$ in (3) can be decomposed into three parts:

$$\hat{H}_{k|i} = H(\theta) + \Psi_{k|i} + O(c^2), \tag{4}$$

where $\Psi_{k|i}$ is a $p \times p$ matrix of terms dependent on $H(\theta)$ and $\Delta_{k|i}$. Specifically,

$$\Psi_{k|i}(H) = \frac{1}{2} H D_{k|i} + \frac{1}{2} D_{k|i}^T H,$$

where $D_{k|i} = \Delta_{k|i}(\Delta_{k|i}^{-1})^T - I_p$ and $I_p$ is the $p \times p$ identity matrix.

Notice that for any value of $H$, $E(\Psi_{k|i}(H)) = 0$. Subtracting both sides of (4) by $\Psi_{k|i}$ and use $\hat{H}_{k|i} - \Psi_{k|i}$ as an estimate of $H(\theta)$, we end up with reduced variance of the Hessian estimate while the expectation of the estimate remains the same. Ultimately, the variance of the estimate of $F(\theta)$ is also reduced. Based on this idea, Spall (2008) introduces a feedback-based method to improve the accuracy of the estimate of $F(\theta)$. The recursive (in $i$) form of the feedback-based form of the estimate of $F(\theta)$, say $\bar{F}'_{M,N}(\theta)$, is

$$\bar{F}'_{M,i}(\theta) = \frac{i-1}{i} \bar{F}'_{M,i-1}(\theta) + \frac{1}{iM} \sum_{k=1}^{M} \left[ \hat{H}_{k|i} - \Psi_{k|i}(\bar{F}'_{M,i-1}(\theta)) \right], \tag{5}$$

where $\bar{F}'_{M,0}(\theta) = 0$. More recent work regarding the feedback-based approach includes Spall (2009), where the feedback ideas are applied to stochastic approximation.

**2.3 Enhancements Through Use of Independent Perturbation per Measurement**

If the $n$ vectors entering each $Z_{\text{pseudo}}(i)$ are mutually independent, the estimation of $F(\theta)$ can be improved by exploiting this independence. In particular, for the basic resampling-based approach, the variance of the elements of the individual Hessian estimates $\hat{H}_{k|i}$ can be reduced by decomposing $\hat{H}_{k|i}$ into a sum of $n$ independent estimates, each corresponding to one of the data vectors. A separate perturbation vector can then be applied to each of the independent estimates, which produces variance reduction in the resulting estimate $\bar{F}_{M,N}(\theta)$. The independent perturbations above reduce the variance of the elements in the estimate of $F(\theta)$ from $O(1/N)$ to $O(1/nN)$.

Similarly, this independent perturbation idea can be applied to the feedback-based approach as well. Besides applying separate perturbation vectors to each of the independent estimates of $\hat{H}_{k|i}$, we also decompose the $\bar{F}'_{M,i-1}(\theta)$ in (5) into a sum of $n$ independent estimates and then apply the $\Psi_{k|i}$ function to individual estimates to gain feedback information to improve the corresponding independent estimates of $\hat{H}_{k|i}$.

## 3. THEORY

The following results are given in Spall (2008) as a theoretical validation for the advantage of the feedback-based approach.

**Lemma**

For some open neighborhood of $\theta^*$, suppose the forth derivative of the log-likelihood function $L''''(\theta)$ exists continuously and that $E\left(\|L''''(\theta)\|^2\right)$ is bounded in magnitude. Furthermore, let $E\left(\|\hat{H}_{k|i}\|^2\right) < \infty$, then for any fixed $M \geq 1$ and all $c$ sufficiently small,

$$E\left(\left\|\bar{F}'_{M,N} - F^* - B(\theta^*)\right\|^2\right) \to 0 \text{ as } N \to \infty,$$

where $B(\theta^*)$ is a bias matrix satisfying $B(\theta^*) = O(c^2)$.

**Theorem**

Suppose that the conditions of the Lemma hold, $p \geq 2$, $E\left(\left\|H(\theta^*)\right\|^2\right) < \infty$, $F^* \geq 0$, and $F^* \neq 0$. Further, suppose that for some $\delta > 0$ and $\delta' > 0$ such that $(1+\delta)^{-1} + (1+\delta')^{-1} = 1$, $E\left(\left\|L''''(\theta)\right\|^{2+2\delta'}\right)$ is uniformly bounded in magnitude for all $\theta$ in an open neighborhood of $\theta^*$, $E\left(\left|1/\Delta_{kj|i}^{2+2\delta}\right|\right) < \infty$. Then the accuracy of $\bar{F}'_{M,N}(\theta)$ is greater than the accuracy of $\bar{F}_{M,N}(\theta)$ in the sense that

$$\lim_{N\to\infty} \frac{E\left[\left\|\bar{F}'_{M,N} - F^*\right\|^2\right]}{E\left[\left\|\bar{F}_{M,N} - F^*\right\|^2\right]} \leq 1 + O(c^2). \tag{6}$$

**Corollary**

Suppose that the conditions of the Theorem hold, $\text{rank}(F^*) \geq 2$, and the elements of $\Delta_{k|i}$ are generated according to the Bernoulli $\pm 1$ distribution. Then, the inequality in (6) is strict.

## 4. NUMERICAL STUDY

In this section, we show the merit of the enhanced Monte Carlo methods over the basic Monte

Carlo resampling method. The performance of the estimation is measured by the relative norm of the deviation matrix: $\|F_{est}(\boldsymbol{\theta}) - F_n(\boldsymbol{\theta})\| / \|F_n(\boldsymbol{\theta})\|$, where the standard spectral norm (the largest singular value) is used, $F_n(\boldsymbol{\theta})$ is the true information matrix and $F_{est}(\boldsymbol{\theta})$ stands for the estimated information matrix via either the basic or the enhanced Monte Carlo approach, as appropriate. For the purpose of comparison, we test under the cases where the true Fisher information is achievable or the exact Hessian matrix is computable, which are not the type of problems we would actually deal with in practice with these estimation methods.

### 4.1 Example 1—Multivariate Normal Distribution in a Signal-Plus-Noise Setting

Suppose that $z_i$'s are independently distributed $N(\boldsymbol{\mu}, \boldsymbol{\Sigma} + \boldsymbol{P}_i)$ for all $i$, where $\boldsymbol{\mu}$ and $\boldsymbol{\Sigma}$ are to be estimated and $\boldsymbol{P}_i$'s are known. This corresponds to a signal-plus-noise setting where the $N(\boldsymbol{\mu}, \boldsymbol{\Sigma})$-distributed signal is observed in the presence of independent $N(\boldsymbol{0}, \boldsymbol{P}_i)$-distributed noise. The varying covariance matrix for the noise may reflect different quality measurements of the signal. This setting arises, for example, in estimating the initial mean vector and covariance matrix in a state-space model from a cross-section of realizations (Shumway, Olsen, and Levy 1981), in estimating parameters for random-coefficient linear models (Sun 1982), or in small area estimating in survey sampling (Ghosha and Rao 1994).

Let us consider the following scenario: $\dim(z_i) = 4$, $n = 30$, and $\boldsymbol{P}_i = \sqrt{i} \boldsymbol{U}^T \boldsymbol{U}$, where $\boldsymbol{U}$ is generated according to a 4×4 matrix of uniform (0, 1) random variables (so $\boldsymbol{P}_i$'s are identical except for the scale factor $\sqrt{i}$). Let $\boldsymbol{\theta}$ represent the unique elements in $\boldsymbol{\mu}$ and $\boldsymbol{\Sigma}$; hence, $p = 4 + 4(4+1)/2 = 14$. So, there are $14(14+1)/2 = 105$ unique terms in $F_n(\boldsymbol{\theta})$ that are to be estimated via the Monte Carlo methods (basic or enhanced approaches). The value of $\boldsymbol{\theta}$ used to generate the data is also used as the value of interest in evaluating $F_n(\boldsymbol{\theta})$. This value corresponds to $\boldsymbol{\mu} = \boldsymbol{0}$ and $\boldsymbol{\Sigma}$ being a matrix with 1's on the diagonal and 0.5's on the off-diagonals. The gradient of the log-likelihood function and the analytical form of the FIM are available in this problem (see Shumway, Olsen, and Levy 1981).

Throughout the study, elements in perturbation $\boldsymbol{\Delta}_{k|i}$ have symmetric Bernoulli ± 1 distribution for all $k$ and $i$; $M = 2$; $c = 0.0001$. In each method, we estimate the Hessian matrix in two different approaches: using the gradient of the log-likelihood function or using the log-likelihood function values only. Results based on 50 independent replications are summarized in Table 1 ($p$-values correspond to $t$-tests of the comparison between the relative norms of the deviation matrices from two approaches).

Table 1: Sample mean value of $\|F_{est}(\boldsymbol{\theta}) - F_n(\boldsymbol{\theta})\| / \|F_n(\boldsymbol{\theta})\|$ with approximate 95% CI shown in brackets. $P$-values based on one-sided $t$-test using 50 independent runs.

| Input Information | Basic Approach | Feedback-based Approach | $p$-value |
|---|---|---|---|
| Gradient Function $N = 40{,}000$ | 0.0104 [0.0096, 0.0111] | 0.0063 [0.0058, 0.0067] | $<10^{-10}$ |
| Log-likelihood Function Only $N = 40{,}000$ | 0.0272 [0.026, 0.0283] | 0.0261 [0.0251, 0.0271] | 0.0016 |
| Log-likelihood Function Only $N = 80{,}000$ | 0.0204 [0.0194, 0.0213] | 0.0191 [0.0184, 0.0198] | $2.52 \times 10^{-5}$ |

Table 1 indicates that there is statistical evidence for the advantage of the feedback-based Monte

Carlo method over the basic Monte Carlo resampling method. The difference between the two methods is more significant when the gradient information of the log-likelihood function is available (row 2) or the number of iterations increases when only likelihood function is available (rows 4).

Keeping all other settings and parameters the same, we now test on the independent perturbation per measurement idea in section 2.3. Table 2 summarizes the simulation results based on 50 independent realizations (*p*-values correspond to *t*-tests of the comparison between the relative norms of the deviation matrices from two approaches).

Table 2: Sample mean value of $\|F_{\text{est}}(\theta) - F_n(\theta)\|/\|F_n(\theta)\|$ when using independent perturbation per measurement. Approximate 95% CIs shown in brackets. *P*-value based on one-sided *t*-test using 50 independent runs.

| Input Information | Indep. Perturbation Alone | Feedback and Indep. Perturbation | *p*-value |
|---|---|---|---|
| Gradient Function $N = 40,000$ | 0.0066 [0.0043, 0.0103] | 0.0062 [0.0044, 0.0097] | $7.622 \times 10^{-9}$ |

Table 2 demonstrates the improvement in estimation accuracy when the sample is independent and separate perturbation is applied to each independent measurement. Specifically, the estimation accuracy is improved by independent perturbation alone (column 2) and is improved even more by the combination of independent perturbation and feedback approach (column 3).

**4.2 Example 2—Mixture Gaussian Distribution**

Mixture Gaussian distribution is of great interest and is popularly used in practical applications (see Wang 2001 and Stein et al. 2002). In this study, we consider a mixture of two univariate normal distributions. Specifically, let $Z = [z_1, z_2, \ldots, z_n]^T$ be an independent and identically distributed sequence with probability density function:

$$f(z, \boldsymbol{\theta}) = \lambda \exp\left(-(z-\mu_1)^2/(2\sigma_1^2)\right)\Big/\sqrt{2\pi\sigma_1^2} + (1-\lambda)\exp\left(-(z-\mu_2)^2/(2\sigma_2^2)\right)\Big/\sqrt{2\pi\sigma_2^2},$$

where $\boldsymbol{\theta} = [\lambda, \mu_1, \sigma_1, \mu_2, \sigma_2]^T$. There are $5(5+1)/2 = 15$ unique terms in $\boldsymbol{F}_n(\boldsymbol{\theta})$ that are to be estimated. The analytical form of the true Fisher information matrix is not attainable in this case. But the closed form of the Hessian matrix is computable (see Boldea and Magnus 2009). We thus approximate the true Fisher information using the sample average of the Hessian matrix over a large number ($10^6$) of independent replications. This should be a fairly good approximation since the decimal accuracy does not vary much as the amount of averaging increases.

In this numerical study, we consider the case where $\boldsymbol{\theta} = [0.2, 0, 1, 4, 9]^T$. As in Example 1, elements in perturbation $\boldsymbol{\Delta}_{k|i}$ have symmetric Bernoulli $\pm 1$ distribution for all $k$ and $i$; $M = 2$; $c = 0.0001$. In each method, we estimate the Hessian matrix in two different approaches: using the gradient of the log-likelihood function or using the log-likelihood function values only. Results based on 50 independent replications are summarized in Table 3 ($p$-values correspond to $t$-tests of the comparison between the relative norms of the deviation matrices from two approaches).

Table 3 indicates that there is statistical evidence for the advantage of the feedback-based Monte Carlo method over the basic Monte Carlo resampling method. The difference between the performances of the two methods is more significant when gradient information of the log-likelihood function is available (row 2) or the number of iterations increases when only likelihood function is available (row 4).

Table 3: Sample mean value of $\|F_{est}(\theta) - F_n(\theta)\|/\|F_n(\theta)\|$ with approximate 95% CI shown in brackets. $P$-values based on one-sided $t$-test using 50 independent runs.

| Input Information | Basic Approach | Feedback-based Approach | $p$-value |
|---|---|---|---|
| Gradient Function $N = 40,000$ | 0.0038 [0.0035, 0.0042] | 0.0013 [0.0011, 0.0015] | $<10^{-10}$ |
| Log-likelihood Function Only $N = 40,000$ | 0.0094 [0.0088, 0.01] | 0.0088 [0.0083, 0.0094] | $2.39 \times 10^{-4}$ |
| Log-likelihood Function Only $N = 80,000$ | 0.0065 [0.006, 0.0069] | 0.0059 [0.0054, 0.0063] | $3.6 \times 10^{-7}$ |

## 5. CONCLUSIONS

This paper demonstrates two enhanced Monte Carlo methods for estimating the Fisher information matrix: feedback-based approach and independent perturbation approach. Numerical examples show that both of these two methods improve the estimation accuracy as compared to the basic Monte Carlo approach.